\RequirePackage{iftex}
\ifLuaHBTeX%
\RequirePackage{luatex85}%
\fi
\documentclass[sigconf,twocolumn,nonacm,natbib=false]{acmart}
\ifLuaHBTeX%
\usepackage{fixme}
\fxsetup{status=draft}
\fi
\usepackage{tikz}
\usepackage{braket}
\usepackage{cleveref}
\usepackage{mathtools}
\setcopyright{none}
\ifLuaHBTeX%
  \usepackage{minted}%
\else
  \usepackage[frozencache,cachedir=.]{minted}%
\fi
\definecolor{codebg}{rgb}{0.95,0.95,0.95}
\setminted{bgcolor=codebg,bgcolorpadding=5pt,linenos=true}
\setmintedinline{bgcolor={},bgcolorpadding=0pt}
\usepackage{pgfplots}
\pgfplotsset{compat=1.18}
\newcommand{\code}[1]{\mintinline{python}{#1}}
\newcommand{\rcode}[1]{\mintinline{rust}{#1}}
\acmConference[EGRAPHS '25]{EGRAPHS 2025}{June 16--20, 2025}{Seoul, South Korea}

\RequirePackage[datamodel=acmdatamodel,style=acmauthoryear]{biblatex}
\newcommand{\R}{\mathbb{R}}

\begin{document}

\title[Detecting Mathematical Optimization Constraints with E-Graphs in JijModeling]{Optimizing Optimizations: Case Study on Detecting Specific Types of Mathematical Optimization Constraints with E-Graphs in JijModeling}

\author{Hiromi Ishii}
\email{h.ishii@j-ij.com}
\orcid{0000-0002-7752-1782}
\author{Taro Shimizu}
\email{t.shimizu@j-ij.com}
\author{Toshiki Teramura}
\email{t.teramura@j-ij.com}
\orcid{0000-0001-9451-9038}
\affiliation{%
  \institution{Jij, Inc.}
  \city{Minato-ku}
  \state{Tokyo}
  \country{Japan}
}

\renewcommand{\shortauthors}{Ishii et al.}
\settopmatter{printfolios=true}
\begin{abstract}
  In solving mathematical optimization problems efficiently, it is crucial to make use of information about specific types of constraints, such as the one-hot or Special-Ordered Set (SOS) constraints. In many cases, exploiting such information gives asymptotically better execution time.
  \emph{JijModeling}~\cite{jijmodeling:pypi,jijmodeling:2025}, an industrial-strength mathematical optimization modeller, achieves this by separating the symbolic representation of an optimization problem from the input data.

  In this paper, we will report a real-world case study on a constraint detection mechanism \emph{modulo the algebraic congruence} using e-graphs, and describe heuristic criteria for designing rewriting systems. We give benchmarking result that shows the performance impact of the constraint detection mechanism.
  We also introduce \texttt{egg\_recursive}~\cite{egg_recursive}, a utility library for writing \texttt{egg}-terms as recursive abstract syntax trees, reducing the burden of writing and maintaining complex terms in S-expressions. 
\end{abstract}

\ccsdesc[500]{Software and its engineering~Domain specific languages}
\ccsdesc[500]{Applied computing~Operations research}

\keywords{e-graphs, mathematical optimization, symbolic processing, Python, Rust, JijModeling}

\maketitle

\section{Introduction\label{sect:intro}}

\emph{Mathematical optimization} is a field of study that deals with finding the ``optimal'' solution for the problem described by a set of constraints and an objective function.
One example is the Travelling Salesman Problem, in which one visits all of $N$ cities exactly once and returns to the starting city, minimizing the total distance travelled.
This can be defined by the following mathematical model (in the quadratic formulation~\cite{10.3389/fphy.2014.00005}):

\begin{alignat*}{2}
  \mathop{\arg\min}_{x_{i,t}}&\quad&& \sum_{i,j,t=0}^{N-1} d_{i,j} x_{i,t}x_{j,(t + 1) \% N} \\
  \text{s.t.} &&&
    \sum_{i=0}^{N-1} x_{i,t} = 1 \; \forall t, \quad
    \sum_{i=0}^{N-1} x_{i,t} = 1 \; \forall i, \\
    &&& x_{i,t} \in \set{0, 1}.
\end{alignat*}
Here, $d_{i,j}$ is the given distance parameter between the $i$-th and $j$-th city, and $x_{i,t}$ is a binary decision variable that is $1$ if the $i$-th city is visited at time $t$. The constraint $\sum_{j=0}^{N-1} x_{i,t} = 1 \; \forall t$ requires that exactly one city is visited at each time step, whereas $\sum_{i=0}^{N-1} x_{i,t} = 1 \; \forall i$ requires that each city is visited exactly once.
These are typical examples of \emph{one-hot constraints} (or \emph{unit-simplex constraints}), i.e.\ constraints that require exactly one of the given binary decision variables to be $1$.
Some solvers can take advantage of such a structure to solve the problem more efficiently, but in most cases we must call the dedicated APIs to define a specific constraint explicitly as a one-hot constraint.
This is because it is impractical to detect such constraints when the input size $N$ gets relatively large.

Another kind of challenge in a constraint detection is that there is more than one way to express the same constraint. For example, users can write one-hot constraints on binary variables $x_i$'s in any of, but not limited to, the following equivalent forms according to their preference:
\[
	\sum_i x_i = 1, \qquad
	\sum_i x_i - 1 = 0, \qquad
	0 = 1 - \sum_i x_i.
\]

Hence, detectors should take algebraic congruences into account. To summarize, we have the following goals in \emph{Constraint Detection Problem}:

\begin{enumerate}
  \item Detect prespecified types of constraints from the given mathematical optimization problem.
  \item Detection mechanism must be able to detect constraints modulo the algebraic congruence.
\end{enumerate}

\emph{JijModeling}~\cite{jijmodeling:pypi,jijmodeling:2025}, a versatile and industrial-strength mathematical optimization modeller, solves these challenges by separating the symbolic representations of optimization problems from the input data and perform pattern matching on them.
JijModeling uses \verb!egg!~\cite{2021-egg} under the hood to handle the algebraic congruence correctly.
To reduce the burden of writing complex rewrite rules and patterns, we also devised a utility library \verb!egg_recursive!~\cite{egg_recursive}, which allows us to write a rule or pattern as a recursive abstract syntax tree.
Compared to the default S-expression-based API, this allows to write more readable and maintainable term rewriting system.

In what follows, we will report a case study about the constraint detection mechanism with egg in JijModeling.
In particular, we will discuss the following points:

\begin{enumerate}
  \item Some heuristic criteria for designing rewrite rules and analysis phase for a constraint detection mechanism.
  \item The design of \verb!egg_recursive! and how it can ease the implementation of multiple rewrite rules.
  \item The runtime impact of the constraint detection mechanism on the solving time of a mathematical optimization problem.
\end{enumerate}

\subsection{Structure of this paper}

This paper is organized as follows.
First in \Cref{sect:overview}, we will give a brief overview of JijModeling and its constraint detection mechanism with some examples.
Then in \Cref{sect:constraint-detection} we describe the architecture of the constraint detection mechanism of JijModeling, including rewrite rules and the example use of our \verb!egg_recursive! crate.
We will also discuss some heuristics for picking the rewrite rules to reduce execution time.
\Cref{sect:benchmarks} shows empirical results on the performance change in solving a mathematical optimization problem using the constraint detection mechanism.
We discuss some future works in \Cref{sect:future} and finally conclude in \Cref{sect:conclusion}.

\section{Overview of JijModeling and Its Constraint Detection\label{sect:overview}}

\begin{listing}[tbp]

\begin{minted}[escapeinside=@@]{python}
import jijmodeling as jm
N = jm.Placeholder("N", dtype=jm.DataType.INTEGER)
d = jm.Placeholder("d", shape=(N,N), ndim=2)
x = jm.BinaryVar("x", shape=(N,N))
i = jm.Element("i", belong_to=N)
j = jm.Element("j", belong_to=N)
t = jm.Element("t", belong_to=N)
prob = jm.Problem(
  "TSP", sense=jm.ProblemSense.MINIMIZE)
prob += jm.Constraint(
  "one_city", x[:,t].sum() == 1, forall=t)
prob += jm.Constraint("one_time", 
  2 * x[i,:].sum() - 1 == 1, forall=i @\label{line:one_time}@
)
prob += jm.sum([i,j,t], 
  d[i,j] * x[i,t] * x[j,(t + 1) % N])
interp = jm.Interpreter({
  'N': 3, 'd': [[0,9,1],[2,0,5],[4,1,0]]
})
inst = interp.eval_problem(prob)
# >>> inst.raw.constraint_hints.one_hot_constraints
# 6
\end{minted}
\caption{JijModeling implementation of TSP\label{code:tsp}}
\end{listing}

Typically, (at least) two tools are involved in applying mathematical optimization to the practical problems:
\begin{description}
	\item[Modeller] is used to describe and encode the real-world problems as (some form of) mathematical equations.
	\item[Solver] solves the encoded problem numerically.
\end{description}
So, the responsibility of a modeller is to provide a convenient way for users to express their problems, convert them into suitable form and feed to solvers.

\emph{JijModeling} is, as its name suggests, classified as a modeller.
Given a description of a mathematical optimization problem and input data, JijModeling compiles them into an OMMX Message~\cite{ommx}, an open-source solver-independent format for mathematical optimization problems.
OMMX comes with adapters for various solvers, and the user can freely pick which solver to use.

What makes JijModeling unique is a \emph{separation} of mathematical expressions and the instance data.
The vast majority of modellers today treat the equations and input data in a mixed form.
In particular, the range of array indices are instantiated in-place right in the equations, which means a reduction operator such as $\sum$ is expanded into a chain of binary additions.
On the other hand, JijModeling uses the dedicated symbolic representation\footnote{JP Patent 7034528} of an optimization problem and stores it separately from the actual input data.
In this way, JijModeling allows a solver-independent description of mathematical optimization problems and can exploit symbolic information to detect specific types of constraints regardless of the actual input data.

A typical JijModeling program encoding the TSP is given in \Cref{code:tsp}\footnote{Currently, we are working on a major update towards JijModeling 2, which will provide more natural syntax without \code{Element}.}.
Note that the \code{one_time} constraint (Line~\ref{line:one_time}) is written in a rather obfuscated form, but JijModeling successfully detects it as a family of one-hot constraints.

The current language of JijModeling consists of two layers:

\begin{description}
  \item[Expressions] incorporating arithmetic operations, reduction operators such as $\sum$ and $\prod$, order-theoretic functions, tensor operations, and boolean expressions.
  \item[Constraint Terms] that take comparison kind ($=$, $\leq$, or $\geq$) and two expressions for left- and right-hand sides.
\end{description}

So our formulation in egg should take these layer distinctions and type information inside expressions into account to maintain the soundness of the system.

\section{Constraint Detection in Action\label{sect:constraint-detection}}

\begin{figure*}[tbhp]
  \begin{center}
    \columnwidth=\linewidth
    \begin{gather}
      a = b \longrightarrow b = a, \quad
      a \leq b \longleftrightarrow b \geq a, \quad
      a + b \lesseqgtr c \longrightarrow a \lesseqgtr c - b, \quad
      a + c \lesseqgtr b + c \longrightarrow a \lesseqgtr b, \quad
      a \cdot c = b \cdot c \longrightarrow a = b \; (\text{if } c \neq 0), \label{eq:constr-orders}\\
      a + b \longrightarrow b + a,\qquad
      (a + b) + c \longleftrightarrow a + (b + c), \qquad 
      \label{eq:arith-1} \\
      \qquad a + 0 \longrightarrow a, \qquad 0 + a \longrightarrow a, \qquad a \longrightarrow a + 0 \; (\text{if } a \in \R), \label{eq:zero-rev} \\
      (-1) \cdot a \longleftrightarrow -a, \qquad a \cdot (-1) \longleftrightarrow -a, \qquad
      a \cdot 0 \longrightarrow 0, \qquad 0 \cdot a \longrightarrow 0, \label{eq:zero-mul} \\
      (a + b) \cdot c \longleftrightarrow a \cdot c + b \cdot c, \qquad
      a \cdot (b + c) \longleftrightarrow a \cdot b + a \cdot c, \label{eq:distr} \\
      a + (-a) \longrightarrow 0, \qquad (-a) + a \longrightarrow 0,\label{eq:conseq}\\
      {(a \cdot b)}^{-1} \longrightarrow a^{-1} \cdot b^{-1} \; (\text{if } a, b \neq 0), \qquad
      a \cdot a^{-1} \longrightarrow 1 \; (\text{if } a \neq 0), \qquad
      {(a^{-1})}^{-1} \longrightarrow a \; (\text{if } a \neq 0), 
      \label{eq:arith-fin}\\
      c \cdot \sum_i a_i \longrightarrow \sum_i c \cdot a_i, \qquad
      \sum_i 0 \longrightarrow 0, \qquad
      {\left(\prod_i a_i\right)}^c \longrightarrow \prod_i a_i^c, \qquad
      \prod_i 1 \longrightarrow 1,
      \label{eq:reduction}\\
      a \land b \longrightarrow b \land a, \qquad (a \land b) \land c \longleftrightarrow a \land (b \land c), \qquad
      a \lor b \longrightarrow b \lor a, \qquad
      (a \lor b) \lor c \longleftrightarrow a \lor (b \lor c), \\
      (a \land b) \lor a \longrightarrow a, \qquad (a \lor b) \land a \longrightarrow a, \qquad
      \lnot (a \land b) \longleftrightarrow \lnot a \lor \lnot b, \qquad
      \lnot (a \lor b) \longleftrightarrow \lnot a \land \lnot b,\label{eq:boolean} \\
      \min(a, b) \longrightarrow \min(b, a), \ldots\label{eq:lattice}
    \end{gather}
  \end{center}
  \caption{Some rewrite rules implemented in JijModeling\label{fig:rules}}
\end{figure*}

As the expressions are stored in symbolic form, the toughest challenge is \emph{matching modulo the congruence}.
One possible alternative is just to use normalization rules and compare normal forms.
But as described in \Cref{sect:overview}, our term language is rather complicated and hence it seems really hard, if not impossible, to define a canonical normal form.
To circumvent this obstacle, we decided to use \verb!egg! for detection without resorting to normalization.

Currently, we are using \verb!egg! solely for the constraint detection.
The overall constraint detection mechanism proceeds as follows:

\begin{enumerate}
  \item Convert a constraint term into an e-graph, one for each constraint\footnote{As of the time of writing, we are computing e-graphs for each constraint separately for the sake of simplicity, avoiding the need to store the constraint ID in e-graphs. This is not essential, so we are planning to try computing one monolithic e-graph containing all constraints.}.
  \item Saturate the e-graphs applying rewrite rules and analysis, independently for each e-graph. Analysis computes the following things:
    \begin{enumerate}
      \item An approximation of the type of the sub-expressions.\label{item:typing}
      \item Constant folding.
    \end{enumerate}
  \item Use \rcode{Pattern}s as many times as needed on each independent e-graph.
\end{enumerate}

In the following sections, we will elaborate on some criteria about the design of an e-graph-based rewriting system for a detection system. These are rather heuristic and based on our experience.
We will also introduce \verb!egg_recursive!~\cite{egg_recursive} and see how it can ease the implementation of multiple rewrite rules.
Developing this library was motivated by the challenges we faced when expressing complex nested patterns using S-expressions, which became unwieldy and error-prone as the complexity of our rule set increased.

\subsection{Heuristics for Designing Rewrite Rules and  Analysis}

\Cref{fig:rules} shows a subset of the rewrite rules implemented in JijModeling.
Rules~\eqref{eq:constr-orders} are for constraint terms, where $\gtreqless$ matches any of ${=}$, ${\leq}$, and $\geq$; the rest are for general expressions.
Rules~\eqref{eq:arith-1}-\eqref{eq:arith-fin} shows some rules of arithmetic operations on floating point numbers.
A distinctive feature of our rules is that we also include some rules for reduction operators, such as $\sum$ and $\prod$.
As the domains of the $\sum$ usually remain unknown until the AST is compiled with the input data, currently we don't include any expansion rules for reduction rules and implements some kind of distributive laws~\eqref{eq:reduction} only.
Besides these, we also have boolean laws~\eqref{eq:boolean}, lattice-theoretic laws~\eqref{eq:lattice} of $\min$ and $\max$, and boolean-valued comparison operators.
Since our goal is to pattern-match modulo some practical variants, the rewrite rules themselves should be sound but not necessarily complete.

In this section, we give some heuristic insights on how to pick the rewrite rules.

\subsubsection{Bidirectionalize rules when pattern-matching}
In \Cref{fig:rules}, there are some (unconditionally) bidirectional rules, such as $(a + b) + c \longleftrightarrow a + (b + c)$ in \eqref{eq:arith-1} or distributive laws in~\eqref{eq:distr}.
This is not needed if one uses e-graphs for program optimization, but the situation is different for the pattern-matching purpose.
The reason is that matching with \texttt{Pattern} on pre-existing e-graphs \emph{does not} update the existing e-graph.
To see the situation, suppose we have a unidirectional rule $(a + b) + c \longrightarrow a + (b + c)$ only and have an e-graph for $a + (b + c)$.
During e-graph computation, this associative law doesn't fire at all and hence there is no e-node for $(a + b) + c$.
Then, if we try to match it against the pattern $(a + b) + c$, it will fail because of this very absence.

On the other hand, we don't have to bidirectionalize self-symmetric rules, such as $a + b \longrightarrow b + a$ or $a \cdot b \to b \cdot a$.
So, for a purpose of pattern-matching, we need to bidirectionalize asymmetric rules to support a wide variety of congruence.

\subsubsection{Use type information for soundness}

There are some rules that are bidirectional but with some side conditions on the reverse direction.
An example of such rules is \eqref{eq:zero-rev}.
Why can't we just make them simply bidirectional, like $a + 0 \longleftrightarrow a$?
The reason is that we are using a term language with \emph{multiple types}.
As mentioned above, our expressions can be of type $\R$, boolean, tensor, or others.
In this situation, we can safely assume that all sub-terms in $a + 0$ are of type $\R$, but on the other hand arbitrary value $a$ is not necessarily of type $\R$.
If we have an unconditional rule $a + 0 \longleftrightarrow a$, then we can apply it to non-$\R$ types, resulting in ill-typed rewritings, e.g.\ $\texttt{True} \longrightarrow \texttt{True} + 0$.
When the language gets complex, such ill-typed rewrite rules can lead to unsound rewriting results, and indeed we were bitten by such ill-typed rules in the past.

In summary: in a language with multiple types, you should recover type information by a side condition when bidirectionalizing the rules which \emph{reduce} the local type information after a rewrite.

To achieve this, you can either:

\begin{enumerate}
  \item First do the type reconstruction on terms, and then proceed to take the congruence closure using such type information, or
  \item Do the type reconstruction in analysis phase, and use it in the rewrite rule.
\end{enumerate}

This is the very reason why we compute the type information in step (\ref{item:typing}), because we did not have a concrete type system at the time of adding detection mechanism.

\subsubsection{Prefer reductive rules to analyses}

Generally, adding rules can result in runtime degradation, but there are some cases where adding a rule can improve performance.
In our experience, adding a rule that reduces the term size after rewriting can improve performance.
An example of such rules is \eqref{eq:conseq}, say $a + (-a) \longrightarrow 0$ and its commutative variant.
Logically, these follow from rules~\eqref{eq:zero-mul}-\eqref{eq:distr} and constant folding as follows:

\begin{alignat*}{2}
  a + (-a) & \longrightarrow 1 \cdot a + (-1) \cdot a &\qquad& (\texttt{mul-one-l}^{-1}, \texttt{mul-neg}^{-1})\\
  & \longrightarrow \left(1 + (-1)\right) \cdot a &\qquad& (\texttt{add-mul-distr}^{-1})\\
  & \mathrel{\phantom{\longrightarrow} \mathllap{\leadsto}} 0 \cdot a &\qquad& (\text{Constant Folding})\\
  & \longrightarrow 0 &\qquad& (\texttt{mul-zero})
\end{alignat*}

Although $a + (-a) \longrightarrow 0$ is logically redundant, adding this rule improves the runtime for computing the saturation. In our experience, it took approximately 5 seconds to reach the fixed-point before adding $a + (-a) \longrightarrow 0$, but it takes less than 1 second after the addition.

In summary: even if a rule is a logical consequence of other rules and analysis, adding such a rule can improve the runtime when it reduces the term size.
This improves runtime especially when deriving the rule needs an additional analysis phase.

\subsection{\texttt{egg\_recursive}: Use Recursive AST to Write Rules Easily}

\begin{listing*}[tbhp]
\begin{minted}[escapeinside=@@]{rust}
let v = |v: &str| DPat::pat_var(v);
let a = || v("a"); let b = || v("b"); let c = || v("c"); let x = || v("x"); let foralls = || v("foralls");
vec![
  rw!("eq-symm"; DPat::eql_cons(a(), b(), foralls()) => DPat::eql_cons(b(), a(), foralls()))
  rw!("le-ge"; DPat::leq_cons(a(), b(), foralls()) => DPat::geq_cons(b(), a(), foralls())),
  rw!("ge-le"; DPat::geq_cons(a(), b(), foralls()) => DPat::leq_cons(b(), a(), foralls())),
  rw!("trans";
      DPat::constraint(Constraint{sense: p(), left: a() + b(), right: c(), forall_list: foralls()}) @\label{line:record}@
      =>
      DPat::constraint(Constraint{sense: p(), left: a(), right: c() - b(), forall_list: foralls()})
  ),
  // ...
  rw!("add-zero-rev"; a() => a() + 0.0f64; if is_of_type(var("?a"), TypeHint::Scalar)),
]
\end{minted}
  \caption{Example of rules written with \texttt{egg\_recursive}\label{code:egg-recursive}}
\end{listing*}

The rules given in \Cref{fig:rules} are just excerpts from our codebase, in which approximately 120 rules are implemented in total.
The default API of \verb!egg! provides an S-expression-based mechanism for specifying the rules, but it is not very convenient for writing complex nested rules.
Furthermore, its \verb!rewrite! macro parses the S-expression only at the \emph{runtime}, which makes the debugging difficult.

To ease this situation, we have developed \verb!egg_recursive!~\cite{egg_recursive}, a utility library for writing rewrite rules in a recursive abstract syntax tree.
Combining with Rust's \verb!std::ops! traits, we can write rules in a more natural way.
\Cref{code:egg-recursive} shows an example of how to write the rules in \verb!egg_recursive!.

The crate is built on top of \verb!egg! and provides a conversion mechanism and custom additional \verb!Searcher! and/or \verb!Appliers!.
The central traits of this crate are as follows:

\begin{itemize}
  \item \rcode{Recursive} trait, abstracting over recursive expressions and can be converted from/to egg term/patterns.
  \item \rcode{IntoLanguageChildren} trait abstracting over a ``view'' types for \rcode{LanguageChildren}.
\end{itemize}

\rcode{Recursive} trait itself provides a standard fold and unfolding abstraction for recursive ASTs.
To be practical, it also provides methods for structural cloning and reference interleaving functions.
\rcode{Language} macro generates the \rcode{Recursive} implementation for a recursively defined enums together with type synonyms for corresponding patterns.

\rcode{IntoLanguageChildren} trait is rather unique to our library.
The aim of this trait is to provide a more human-readable way of writing $N$-ary AST nodes without memorizing the order of the arguments.
Line~\ref{line:record} in \Cref{code:egg-recursive} shows an example usage of this record-based feature. There, \rcode{Constraint} is an instance of \rcode{IntoLanguageChildren} and the user can use labels like \rcode{sense}, \rcode{left}, \rcode{right}, and \rcode{forall_list} to refer to the corresponding sub-terms without recalling the fixed order.

The following code gives an example of how to define a view type and recursive AST with the derive macros.
\begin{minted}{rust}
#[derive(Debug, Clone, LanguageChildren)]
pub struct IfThenElse<T> {
  pub cond: T, pub then: T, pub else_: T,
}
#[derive(Debug, Clone,  Language)]
pub enum Arith {
  Num(i32),
  Neg(Box<Self>), 
  Add([Box<Self>; 2]), 
  Mul([Box<Self>; 2]),
  IfThenElse(IfThenElse<Box<Self>>),
}  
\end{minted}

\begin{listing}[tbhp]
\begin{minted}{rust}
use DetectorTermPat as DPat;
use DPat::pat_var as var;
DPat::eql_cons(
    DPat::sum(ReductionArgs {
        index: var("index"),
        condition: var("cond"),
        operand: ast::DecisionVar {
            name: var("operand"),
            shape: DPat::list(vec![]),
            kind: DecisionVarKind::Binary.into(),
        }
        .into(),
    }),
    1.0f64.into(),
    var("foralls"),
)
\end{minted}
\caption{Recursive pattern for one-hot constraints\label{code:one-hot}}
\end{listing}

As mentioned above, \rcode{Language} macro also generates the recursive wrapper type to be used as a drop-in replacement of \rcode{Pattern}.
\Cref{code:one-hot} gives a recursive representation of a pattern matching any one-hot constraint, which is a simplified version of the one used in our actual detection mechanism.

\section{Runtime Impact of Constraint Detection\label{sect:benchmarks}}

We have discussed the importance of constraint detection so far. But how much does it actually have an effect?
In this section, we present the empirical results of the performance change in solving a mathematical optimization problem using the constraint detection mechanism.

To measure the impact on performance, we use the following plant placement problem, which is a variant of the problem from~\cite[\S~4.9]{santos2020mixed}:

\begin{alignat*}{2}
  \mathop{\arg \min}_{\delta_i, c_i, s_{ij}} & \sum_{i, j} s_{ij} \left\|\boldsymbol{p}_{p,i} - \boldsymbol{p}_{c,j} \right\| + \sum_i c_i\\
  \text{s.t.} &
    \sum_{\substack{i \leq N \\ x_i < 50}} \delta_i \leq 1,
    &&\sum_{\substack{i \leq N \\ x_i \geq 50}} \delta_i \leq 1,
    \tag{\ensuremath{\star}}\label{eq:sos1-binary}\\
      &\delta_i \in \set{0, 1}      && \forall i \leq N \tag{\ensuremath{\mathord{\star}\mathord{\star}}}\label{eq:delta-bin}\\
      &0 \leq c_i \leq C_i \delta_i,&& \forall i \leq N\tag{\ensuremath{\mathord{\star}\mathord{\star}\mathord{\star}}}\label{eq:xi-sos1}\\
      &\sum_{i \leq N} s_{ij} = d_j,&& \forall j \leq M\\
      &\sum_{j \leq M} s_{ij} = c_i,&& \forall i \leq N
\end{alignat*}

In short, the problem is to pick at most one plant for each of the east ($x_i < 50$) and west ($x_i \geq 50$) areas, and to assign the amount of each product from each plant to each customer, minimizing the total cost.
The important point is that this problem includes so-called \emph{Special-Ordered Set constraints of type 1} (SOS1), which can be efficiently solved by many mixed integer programming solvers.
An SOS1 constraint demands a list of non-negative decision variables to have \emph{at most one} non-zero value.
In the definition above, constraints~\eqref{eq:sos1-binary} to~\eqref{eq:xi-sos1} jointly specify that the two sets of real variables $\Set{c_i \in [0, C_i] | x_i < 50}$ and $\Set{c_i | x_i \geq 50}$ are subject to SOS1 constraints respectively.
We define this problem in JijModeling 1.12.3 and solve it with the PySCIPOpt 1.8.1~\cite{pyscipopt} with Python 3.10.15.
The entire benchmark code is available on GitHub~\cite{bench}.

\begin{figure}[tbp]
  \begin{center}
    \begin{tikzpicture}
      \begin{axis}[width=0.9\columnwidth,legend pos=north west,xlabel=$N$,ylabel={CPU Time $[\mathrm{sec}]$},xtick distance=6]
        \addplot[black,mark=*,sharp plot] table[x=N,y=without-detection,header=has colnames,col sep=comma] {sos1-bench.csv};
        \addplot[red,mark=+,sharp plot] table[x=N,y=with-detection,header=has colnames,col sep=comma] {sos1-bench.csv};
        \legend{Without Detection,With Detection}
      \end{axis}
    \end{tikzpicture}
  \end{center}
  \caption{Runtime of solving the plant placement problem, with and without SOS1 detection\label{fig:plant-bench}}
\end{figure}
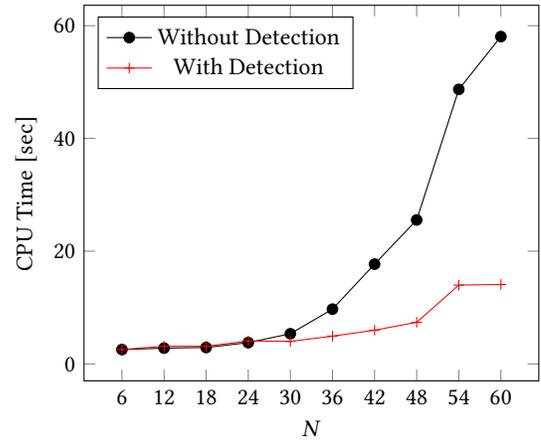

\Cref{fig:plant-bench} shows the benchmark result of the plant placement problem with and without constraint detection. The benchmark is taken with the \texttt{pytest-benchmark} framework~\cite{pytest-benchmark} on a MacBook Pro 2024 with Apple M3 Chip (8 cores) and 16 GB of RAM.
The plot shows that the problem can be solved drastically faster with SOS1 detection.

\section{Future Works\label{sect:future}}

Of course, our system is not perfect and there is much room for improvement.
We will discuss some issues and possible future works in this section.

First, we are planning to use \verb!egglog!~\cite{egglog}, the successor to the \verb!egg! combined with the Datalog language.
This should be useful when one wants to write composite detection rules.
For example, an SOS1 constraint on \emph{general} decision variables is usually expressed as a combination of an SOS1 constraint on \emph{binary variables} and upper-bound constraints on the decision variables multiplied with binaries.
Currently, to detect this form of SOS1, we are matching against SOS1 constraints on binary variables, and use that information again to detect the general cases.
This requires multiple instances of manual pattern-matching, and we have to write such conditionals manually outside the rules.
As this kind of condition can easily be expressed as (cut-free) Horn clauses, we believe that egglog should ease the situation further.

Another issue is the treatment of \emph{bound variables}.
Currently, JijModeling treats bound variables in a rather ad-hoc way requiring users to define bound variables separately from binders.
This accidentally eases the pattern-matching on expressions incorporating binders, e.g.\ $\sum$ or $\prod$ in our case for the time being.
But considered as a language, such a formulation is error-prone.
To fix this, we are currently working on proper treatment of bound variables using either the locally nameless~\cite{Chargueraud:2012aa} or the higher-order abstract syntax~\cite{Pfenning:1988aa} approach.
This change, on the other hand, imposes another challenge in constraint detection, as the interaction of $\beta$-expansion and e-graphs is not well understood and indeed an active research area these days.

Finally, we need a more systematic mechanism for matching on \emph{variadic operators}.
Currently, we are expressing additions and/or multiplications in a binary way, i.e.\ $a + b + c$ is expressed as $((a + b) + c)$.
Since they are commutative and associative (up to floating-point error), it might be better to express them as a list or bag of summands and make a direct pattern-matching on them.
This could perhaps eliminate the necessity of applying associative and commutative laws and can help improve performance.
To achieve this, we need a way to express such terms and a handy way of making pattern-matching on them.

\section{Conclusion\label{sect:conclusion}}

We have seen that, in solving an optimization problem, it is crucial to detect the usages of specific types of constraints \emph{modulo algebraic congruence}.
Such information is actually vital to speed up the solving process of an optimization problem by invoking dedicated algorithms implemented in the solvers.
To solve this problem, we have successfully applied e-graph-based equality saturation to realize a fast and clever constraint detection mechanism in JijModeling.

While implementing it, we have learnt the following heuristic lessons for designing a rewriting system:
\begin{enumerate}
  \item Bidirectionalize rules when you want to pattern-match.
  \item To avoid unsound rewriting due to ill-typed rules, add side-conditions to recover type information when writing the inverse rule of a rule that reduces the type information.
  \item Prefer reductive rules to analyses to gain performance improvement.
\end{enumerate}

We also developed \verb!egg_recursive!, a utility library for writing rewrite rules in a recursive abstract syntax tree.
With this, we can write egg rewrite rules in a more natural and confident way.

Although we focused on the particular application of a constraint detection, but methods and lessons we discussed in this paper can be applied to other applications involving pattern-matching modulo congruence.

\begin{acks}
  Many thanks to our colleagues: Iago Almeida tunes up JijModeling making the benchmark result prominent; Hiromichi Matsuyama also gave us many useful comments on the draft.

  A part of this work was performed for Council for Science, Technology and Innovation (CSTI), Cross-ministerial Strategic Innovation Promotion Program (SIP), ``Promoting the application of advanced quantum technology platforms to social issues'' (Funding agency: QST).
\end{acks}

\printbibliography{}

@online{bench,
	author = {{Jij, Inc.}},
	title = {Jij-Inc/sos1-detection-benchmarks at 4e9fe48da5795694aa0010f429ea8ec944860e9b},
	url = {https://github.com/Jij-Inc/sos1-detection-benchmarks/tree/4e9fe48da5795694aa0010f429ea8ec944860e9b},
	urldate = {2025-04-16},
	year = {2025}}

@article{10.3389/fphy.2014.00005,
	author = {Lucas, Andrew},
	doi = {10.3389/fphy.2014.00005},
	issn = {2296-424X},
	journaltitle = {Frontiers in Physics},
	title = {{Ising} formulations of many {NP} problems},
	url = {https://www.frontiersin.org/journals/physics/articles/10.3389/fphy.2014.00005},
	volume = {Volume 2 - 2014},
	year = {2014}}

@article{santos2020mixed,
	author = {Santos, Haroldo G and Toffolo, T},
	journal = {COINOR Computational Infrastructure for Operations Research},
	title = {Mixed integer linear programming with {Python}},
	year = {2020}}

@inproceedings{Pfenning:1988aa,
	address = {New York, NY, USA},
	author = {Pfenning, F. and Elliott, C.},
	booktitle = {Proceedings of the ACM SIGPLAN 1988 Conference on Programming Language Design and Implementation},
	doi = {10.1145/53990.54010},
	isbn = {0897912691},
	location = {Atlanta, Georgia, USA},
	numpages = {10},
	pages = {199--208},
	publisher = {Association for Computing Machinery},
	series = {PLDI '88},
	title = {Higher-order abstract syntax},
	url = {https://doi.org/10.1145/53990.54010},
	year = {1988}}

@article{Chargueraud:2012aa,
	author = {Chargu{\'e}raud, Arthur},
	date = {2012/10/01},
	doi = {10.1007/s10817-011-9225-2},
	id = {Chargu{\'e}raud2012},
	isbn = {1573-0670},
	journal = {Journal of Automated Reasoning},
	number = {3},
	pages = {363--408},
	title = {The Locally Nameless Representation},
	url = {https://doi.org/10.1007/s10817-011-9225-2},
	volume = {49},
	year = {2012}}

@article{egglog,
	address = {New York, NY, USA},
	articleno = {125},
	author = {Zhang, Yihong and Wang, Yisu Remy and Flatt, Oliver and Cao, David and Zucker, Philip and Rosenthal, Eli and Tatlock, Zachary and Willsey, Max},
	doi = {10.1145/3591239},
	issue_date = {June 2023},
	journal = {Proc. ACM Program. Lang.},
	month = {jun},
	number = {PLDI},
	numpages = {25},
	publisher = {Association for Computing Machinery},
	title = {Better Together: Unifying Datalog and Equality Saturation},
	url = {https://doi.org/10.1145/3591239},
	volume = {7},
	year = {2023}}

@online{egg_recursive,
	author = {{Jij, Inc.}},
	title = {egg\_recursive, an S-expression-free alternative interface to egg},
	url = {https://crates.io/crates/egg_recursive},
	urldate = {2025-03-31},
	year = {2024}}

@online{ommx,
	author = {{Jij, Inc.}},
	title = {What is OMMX? -- OMMX},
	url = {https://jij-inc.github.io/ommx/en/introduction.html},
	urldate = {2025-03-31},
	year = {2025}}

@article{2021-egg,
	address = {New York, NY, USA},
	articleno = {23},
	author = {Willsey, Max and Nandi, Chandrakana and Wang, Yisu Remy and Flatt, Oliver and Tatlock, Zachary and Panchekha, Pavel},
	doi = {10.1145/3434304},
	issue_date = {January 2021},
	journal = {Proc. ACM Program. Lang.},
	month = jan,
	number = {POPL},
	numpages = {29},
	publisher = {Association for Computing Machinery},
	title = {egg: Fast and Extensible Equality Saturation},
	url = {https://doi.org/10.1145/3434304},
	volume = {5},
	year = {2021}}

@online{jijmodeling:pypi,
	author = {{Jij, Inc.}},
	title = {jijmodeling · {PyPI}},
	url = {https://pypi.org/project/jijmodeling},
	urldate = {2025-01-29},
	year = {2025}}

@online{jijmodeling:2025,
	author = {{Jij, Inc.}},
	title = {What is {JijModeling}?},
	url = {https://jij-inc.github.io/JijModeling-Tutorials/en/introduction.html},
	urldate = {2025-03-25},
	year = {2023}}

@incollection{pyscipopt,
	author = {Stephen Maher and Matthias Miltenberger and Jo{\~{a}}o Pedro Pedroso and Daniel Rehfeldt and Robert Schwarz and Felipe Serrano},
	booktitle = {Mathematical Software {\textendash} {ICMS} 2016},
	doi = {10.1007/978-3-319-42432-3_37},
	pages = {301--307},
	publisher = {Springer International Publishing},
	title = {{PySCIPOpt}: Mathematical Programming in {Python} with the {SCIP} Optimization Suite},
	year = {2016}}

@online{pytest-benchmark,
	author = {Ionel Cristian M{\u a}rieș},
	title = {pytest-benchmark 5.1.0 documentation},
	url = {https://pytest-benchmark.readthedocs.io/en/latest},
	urldate = {2025-01-30},
	year = {2024}}
\end{document}